\documentclass[sort&compress,final,3p,times,twocolumn,fixfloat]{elsarticle}

\usepackage{graphicx}
\usepackage{amssymb} 
\usepackage{amsmath} 
\usepackage{times}
\usepackage{tabularx}

\usepackage{txfonts}
\usepackage{lineno}

\journal{arXiv: SLAC-PUB-17253}

 \usepackage{ifpdf}
 \ifpdf
 \usepackage[pdftex]{hyperref}
 \else
 \usepackage[hypertex]{hyperref}
 \fi

 \hypersetup{
   pdftitle={},%
   pdfauthor={},%
   pdfsubject={},%
   pdfkeywords={},%
   pdfstartview={},%
   bookmarksopen=true, breaklinks=true, debug=true, %
   colorlinks=true, linkcolor=red, citecolor=blue, urlcolor=blue
 }

\usepackage{float} 
\usepackage[caption=false]{subfig} 

\newcount\savefnused
\newcount\savefndone

\newcommand{\savefootnote}[2][\empty]
{\ifx\empty#1\footnotemark\else\footnotemark[#1]\fi
 \global\advance\savefnused by 1
 \expandafter\xdef\csname savefnmark\the\savefnused\endcsname{\thefootnote}%
 \expandafter\xdef\csname savefntext\the\savefnused\endcsname{#2}%
}
\newcommand{\flushfootnote}{\loop\ifnum\savefndone<\savefnused
  \global\advance\savefndone by 1
  \footnotetext[\csname savefnmark\the\savefndone\endcsname]%
    {\csname savefntext\the\savefndone\endcsname}%
  \global\expandafter\let\csname savefnmark\the\savefndone\endcsname\relax
  \global\expandafter\let\csname savefntext\the\savefndone\endcsname\relax
\repeat}

\usepackage{multirow,bigdelim}

\usepackage{textcomp}
\usepackage{comment}

\usepackage{lineno}

\usepackage{float} 
\usepackage{subfig} 
\usepackage{amsmath} %
\usepackage{amstext}

\usepackage{graphicx}%
\usepackage{tabularx,ragged2e}%

\newcolumntype{Y}{>{\centering\arraybackslash}X}

\def\eg{{\it e.g.}}

\def\GeV{{\rm GeV}}
\def\GeV2{{\rm GeV}^2}

\newcommand{\cf}[1]{{Fig.~\ref{#1}}}

\begin{document} 
\begin{frontmatter}

\title{The gluon and charm content of the deuteron}
\author[a]{Stanley J. Brodsky}
\author[a]{Kelly Yu-Ju Chiu}
\author[b]{Jean-Philippe Lansberg}
\author[b,c]{Nodoka Yamanaka}
\address[a]{SLAC National Accelerator Laboratory, Stanford University, 
 Stanford, CA 94309, USA}
\address[b]{IPNO, CNRS-IN2P3, Univ. Paris-Sud, Universit\'e Paris-Saclay, 
91406 Orsay Cedex, France}
\address[c]{iTHES Research Group, RIKEN, Wako, Saitama 351-0198, Japan}

\begin{abstract}
{\small
We evaluate the frame-independent gluon and charm parton-distribution functions (PDFs) of the deuteron utilizing light-front quantization and the impulse approximation. 
We use a nuclear wave function obtained from solving the nonrelativistic Schr\"{o}dinger equation with the realistic Argonne v18 nuclear force,  which we fold with the proton PDF. 
The predicted gluon distribution in the deuteron (per nucleon) is a few percent smaller than  that of the proton in the domain $x_{bj} = {Q^2\over 2 p_N \cdot q}  \sim 0.4$, whereas it is strongly enhanced for $x_{bj}$ larger than 0.6. 
We discuss the applicability of our analysis and comment on how to extend it to the kinematic limit $x_{bj} \to 2$. 
We also analyze the charm distribution of the deuteron within the same approach by considering both the perturbatively and non-perturbatively generated (intrinsic) charm contributions.
In particular, we note that the intrinsic-charm content in the deuteron will be enhanced due to 6-quark ``hidden-color" QCD configurations.
}
\end{abstract}

%
\end{frontmatter}
%

\section{Introduction}

A primary challenge in nuclear physics is to study the structure and dynamics of nuclei from first principles in terms of the fundamental quark and gluon degrees of freedom of quantum chromodynamics (QCD).
The conventional  description of nuclear many-body systems,  where nucleons are treated as elementary particles with phenomenological potentials, can be justified  in the nonrelativistic domain \cite{Wiringa:1994wb,vanKolck:1994yi,Pieper:2001ap,Kamada:2001tv,Carlson:2014vla,Reinert:2017usi}.    
However, in the short-distance, high-momentum-transfer region, quark and gluon fields play an essential role in describing nuclear systems, and non-nucleonic phenomena, such as QCD ``hidden-color degrees" of freedom \cite{Brodsky:1983vf,Brodsky:1985gt,Brodsky:1985gs,Bashkanov:2013cla}, become relevant. 
For example,  the six-quark Fock state of the deuteron has five different SU(3) color-singlet contributions, only one of which projects to the standard proton and neutron three-quark clusters. 
The leading-twist shadowing \cite{Mueller:1985wy,Brodsky:1989qz,Piller:1999wx,Armesto:2006ph,Frankfurt:2016qca} of nuclear parton distributions at small $x_{bj}$  in the Gribov-Glauber theory is due to the destructive interference of two-step and one-step amplitudes, where the two-step amplitude depends on diffractive deep inelastic scattering (DDIS) $\ell N \to \ell' N' X$, leaving the struck nucleon intact.
The study of the quark and gluon structure of nuclei thus illuminates the intersection between the nuclear and particle physics. 

The quark and gluon distributions of nuclei also play an important role in high-energy astrophysics \cite{Enberg:2008te,Bhattacharya:2015jpa}. 
An accurate knowledge of nuclear parton distributions is essential in many physics fields \cite{Arneodo:1992wf}.
For example, the gluonic content of light nuclei is important in understanding the production of antiprotons in interstellar reactions.
The charm-quark distribution function in nuclei at high $x_{bj}$ can significantly change the predictions of the spectrum of cosmic neutrinos and is thus important to interpret the background of ultra-high-energy neutrinos which contribute to the IceCube experimental data~\cite{Aartsen:2014gkd,Aartsen:2016xlq} in the high-$x_F$ domain~\cite{Halzen:2016thi,Laha:2016dri,Giannini:2018utr}.
Furthermore, the parton-distribution function (PDF) for nuclei is the initial condition controlling the dynamics of the possible formation and thermalization of the quark-gluon plasma (see \eg~\cite{Andronic:2015wma}).

Collider experiments typically probe  proton and nuclear  PDFs in the region of small  $x_{bj} = {Q^2\over 2 p_N \cdot q}$ (see~\cite{Zenaiev:2015rfa,Gauld:2016kpd,Kusina:2017gkz,Lansberg:2016deg} for recent works showing the relevance of LHC heavy-flavor data to determine the gluon content of the nuclei at small $x_{bj}$).
In contrast, fixed-target experiments can unveil the PDF over the full range of $x_{bj}$ up to unity by taking advantage of the asymmetry of the experimental apparatus and the kinematics.
New fixed-target experiments using the beams of the LHC are currently investigated (see the works of the AFTER@LHC study group~\cite{Brodsky:2012vg,Trzeciak:2017csa,Kikola:2017hnp,Massacrier:2015qba,Lansberg:2012kf}) following the very positive outcome of the data taking of the SMOG@LHCb system~\cite{Maurice:2017iom,Anderlini:2017mom}.
In fixed-target experiments, one also has the advantage that the parton distributions of a large variety of nuclei, both polarized and unpolarized, can be measured.
It is thus an important theoretical task to predict the gluon and heavy-quark distributions of nuclei.

We will focus on the deuteron, which is the simplest many-nucleon system, and thus can be evaluated with high accuracy in nuclear physics.
It is therefore an excellent system where  nuclear effects \cite{Brodsky:1983vf,Brodsky:1985gs,Brodsky:1976mn,Berlad:1979tc,Bodek:1980ar,Bodek:1981wr,Aubert:1983xm,Frankfurt:1988nt,Whitlow:1991uw,Merabet:1993du,Kulagin:1994fz,Melnitchouk:1996vp,Hirai:2001np,Sargsian:2002wc,Chen:2004zx,Arrington:2003nt,Kulagin:2004ie,Hirai:2007sx,Accardi:2009br,Weinstein:2010rt,Accardi:2011fa,Ethier:2013hna,Ehlers:2014jpa,Hen:2016kwk,Chen:2016bde,Alekhin:2017fpf,Winter:2017bfs} can be studied.
In addition, a careful study of the structure of the deuteron may provide accurate information on the quark and gluon structure of the neutron \cite{Melnitchouk:1995fc,Afnan:2000uh,Afnan:2003vh}. 
In particular, the gluon PDF of the neutron is of interest.
The PDF of the deuteron near the maximal fraction $x_{bj} = 2$ (we use this definition in this work) can be constrained by  perturbative QCD, since it is the dual of the deuteron form factor at high-momentum transfer $Q^2$~\cite{Brodsky:1973kr,Radyushkin:2009wx}.
In this work, we will mostly be interested in the region of $x_{bj} \sim 1$, a domain which AFTER@LHC can access.

As a first study, we have calculated the gluon PDF in the deuteron within the impulse approximation which  gives the leading contribution at $x_{bj} <1$. 
To do so, we have solved the Schr\"{o}dinger equation of the two-nucleon system with a phenomenological nuclear potential \cite{Wiringa:1994wb} using the Gaussian expansion method \cite{Hiyama:2003cu}.
We have then derived the boost-invariant light-front wave function \cite{Brodsky:1997de,Lepage:1980fj} of the nucleus and convoluted it with  the gluon distribution of the nucleon in order to obtain the gluon distribution of the deuteron. 
The complications of boosting an instant-form nucleon wavefunction to nonzero momentum are discussed in Ref.~\cite{Brodsky:1968ea}.

This paper is organized as follows.
In the next section, we calculate the gluon PDF of the deuteron through the procedure mentioned above.
In Section \ref{sec:result}, we discuss the applicability of the impulse approximation and show our result.
We also extend our discussion to illuminate the intrinsic heavy-quark contribution to the deuteron charm-quark distribution (Section \ref{sec:charm}).
A summary is presented in the final section.

\section{Derivation of the gluon PDF of the deuteron}

\subsection{Deuteron wave function}

Let us now explain how we convolute the gluon PDF of the nucleon by the deuteron wave function in the impulse approximation [see \cf{fig:pdf_operator} (a)].
The impulse approximation is the leading contribution in the chiral effective field theory ($\chi$EFT) \cite{Chen:2004zx,Chen:2016bde}.
We will show later that the two-nucleon contribution [\cf{fig:pdf_operator} (b)] is subleading in the nucleon velocity expansion.
These arguments lead us to consider a nonrelativistic framework.

\begin{figure}[htb]
\begin{center}
\includegraphics[width=8.5cm]{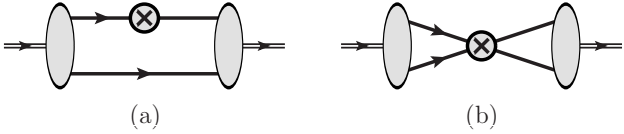}
\caption{\label{fig:pdf_operator}
Schematic representation of the PDF of the deuteron.
The solid and double lines denote the nucleon and the deuteron, respectively, and the cross indicates the PDF operator.
There are two distinct contributions:
(a) one-nucleon operator, working in the impulse approximation, 
(b) two-nucleon operator, relevant in high-momentum exchange.
}
\end{center}
\end{figure}

We first calculate the wave function of the deuteron, given by the bound-state solution of the nonrelativistic two-nucleon Schr\"{o}dinger equation with the Argonne $v18$ potential \cite{Wiringa:1994wb} as the nuclear force.
To solve the equation, we use the Gaussian expansion method \cite{Hiyama:2003cu}, where an accurate solution is provided as a superposition of Gaussians with geometric series of ranges.
The Gaussian basis is given by
\begin{eqnarray}
\Phi_{ n l  m } ( {\mathbf r} )
&=&
N_{nl}
r^l e^{-\nu_n r^2}
Y_{lm} (\hat{r} )
,
\label{eq:gaussian_basis}
\end{eqnarray}
where $N_{nl}$ is the normalization constant of the Gaussian basis, $\hat{r}$ the unit vector of the relative coordinate ${\mathbf r}$, and $\nu_n = \frac{1}{r_n^2} = \frac{1}{r_1 a^{n-1}}$ ($n = 1 ,\cdots , n_{\rm max}$). 
We have taken $n_{\rm max}=12$ Gaussians with $r_1 = 0.1$ fm and the common ratio $a$ so that $r_{12} = 10$ fm.
Note that the nuclear force has a strong tensor force which may change the orbital angular momentum by two units, so the $S$-wave and $D$-wave states are relevant. 
The deuteron state is thus given by
\begin{align}
&| \, ^2{\rm H} , m_j \rangle
=
\sum_n
c^{(s)}_{n}
N_{n0}
e^{-\nu_n r^2}
Y_{00} (\hat{r} )
\chi_{1,m_j}+
\nonumber\
\\
&\sum_{n'} 
c^{(d)}_{n'}
N_{n' 2}
r^2
e^{-\nu_{n'} r^2}
\!\!\sum_{m_l , m_s}\!
f_{m_l m_s m_j} 
Y_{2m_l} (\hat{r} )
\chi_{1,m_s},
\label{eq:deuteron_gaussian}
\end{align}
where $\chi_{1 , m_s} \equiv |\, s = 1 , m_s \rangle$, and $f_{m_l m_s m_j} \equiv \langle l\!=\!2 , m_l , s\!=\!1 , m_s \,|\, j = 1, m_j \rangle$.

To solve the Schr\"{o}dinger equation, we have to diagonalize the Hamiltonian matrix together with the  nuclear norm matrix which involves the information of the overlap between Gaussian basis functions.
This is a generalized eigenvalue problem (For details, see Section 2.1 of Ref. \cite{Hiyama:2003cu}).
By diagonalizing the Hamiltonian, we obtain the wave function shown in \cf{fig:deuteron_wave_function}, which has a dominant $S$-wave component and a $D$-wave component representing 6\% of the total probability.

\begin{figure}[htb]
\begin{center}
\includegraphics[width=8.5cm]{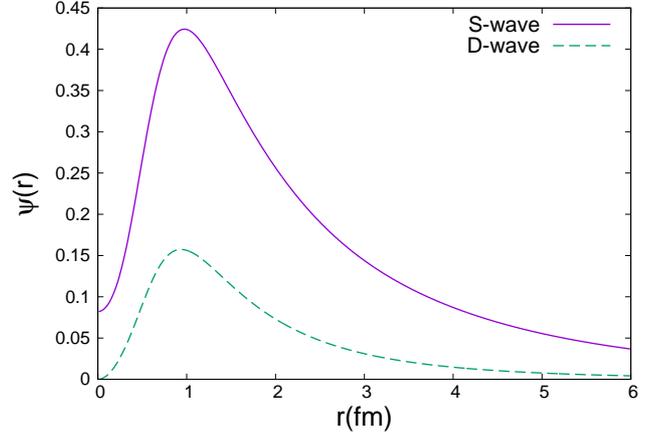}
\caption{\label{fig:deuteron_wave_function}
Radial component (spherical coordinate) of the deuteron wave function.
}
\end{center}
\end{figure}

In our framework, the wave function is given as a superposition of Gaussians, so further transformations can analytically be performed.
We then Fourier transform it and project the wave function onto the $z$-axis.
After some manipulations, we obtain the following expression for the wave function of the unpolarized deuteron expressed in terms of the momentum in the $z$-axis $p_z$:
\begin{eqnarray}
P(p_z)
=
\sum_{n}\sum_{n'} \frac{c_n^{(s)} c_{n'}^{(s)} N_{n 0} N_{n' 0} e^{-\frac{1}{4} \left( \frac{1}{\nu_n} -\frac{1}{\nu_{n'}} \right) p_z^2 }}{8 \sqrt{\nu_n \nu_{n'} } (\nu_n + \nu_{n'}) }&&
\nonumber\\
+
\sum_{m}\sum_{m'} \frac{c_m^{(d)} c_{m'}^{(d)} N_{m 2} N_{m' 2} e^{-\frac{1}{4} \left( \frac{1}{\nu_m} -\frac{1}{\nu_{m'}} \right) p_z^2 }}{32 \sqrt{\nu_m \nu_{m'} } (\nu_m + \nu_{m'}) }&&
\nonumber\\
\hspace{1em} \times
\Biggl\{
\frac{8 }{ (\nu_{m} +\nu_{m'})^2}
+\frac{2 p_z^2 }{\nu_{m} \nu_{m'}(\nu_{m} +\nu_{m'})}
+\frac{p_z^4}{4 \nu_{m}^2 \nu_{m'}^2}  
\Biggr\}&&
.
\label{eq:p(p_z)}
\end{eqnarray}
Let us note that the cross-terms between $c_n^{(s)}$ and $c_{n'}^{(d)}$ cancel.

\begin{figure}[htb]
\begin{center}
\includegraphics[width=8.5cm]{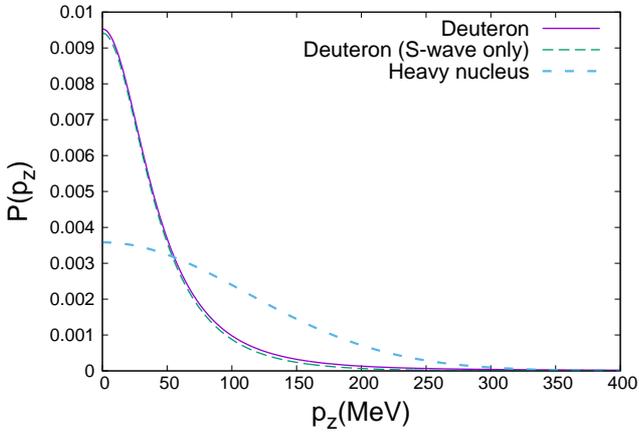}
\caption{\label{fig:Momentum_z-axis_distribution}
Momentum $z$-axis component of the deuteron wave function.
The data for a typical nucleus with a Fermi energy $\epsilon_F = 33$ MeV are also shown for comparison (labeled as ``Heavy nucleus'').
}
\end{center}
\end{figure}

The corresponding probability distribution is shown in \cf{fig:Momentum_z-axis_distribution}. 
The distribution of the nucleon momentum is centered at $p_z=0$, and the standard deviation is close to 50 MeV.
This is due to the kinetic energy of the nucleon (about 20 MeV), which is the bound-state effect of the nuclear force.
Figure \ref{fig:Momentum_z-axis_distribution} also displays the contribution from the $S$-wave, which is nearly identical to the total result.

In \cf{fig:Momentum_z-axis_distribution}, we also show the momentum distribution of the nucleon inside a typical heavy nucleus with the Fermi energy $\epsilon_F \equiv \frac{ p_F^2}{2m_N} = 33$ MeV.
The smearing of the momentum distribution is given by \cite{Merabet:1993du}
\begin{equation}
P(p_z ) 
=
\frac{1}{\sqrt{2\pi \gamma_F}} \exp \Bigl( -\frac{p_z^2}{2\gamma_F} \Bigr)
,
\end{equation}
where $\gamma_F = \frac{1}{5} p_F^2$.
One sees that the momentum distribution of the deuteron is narrower than that of a typical heavy nucleus. 

\subsection{Light-front momentum fraction}

We now calculate the light-front momentum distribution of the nucleon in the deuteron.
Note that the procedure to obtain a wave function in the light-front frame from the instant-form is not unique. 
In this work, we follow the recipe of Ref. \cite{Terentev:1976jk} (see also \cite{Hoyer:1999xe,Hufner:2000jb,Kopeliovich:2001ee,Merabet:1993du}) giving the wave function in the light-front frame as
\begin{equation}
\psi ({\mathbf p}_\perp , z)
=
\sqrt{
\frac{\partial p_z ({\mathbf p}_\perp, z)}{\partial z}}
\psi ({\mathbf p}_\perp, p_z)
,
\end{equation}
where $p_z = (z -1) \sqrt{\frac{ m_N^2 +{\mathbf p}_\perp^2}{ z(2-z)}}$.
The momentum fraction of the nucleon in the deuteron $z$ is defined in the interval $0 \le z \le  2$.
This can consistently be derived using $z$ defined by
\begin{equation}
z \equiv 
A \frac{p_N^+}{p_A^+} = \frac{A}{m_A} \Biggl[ \sqrt{m_N^2 +p_z^2 + {\mathbf p}^2_\perp } + p_z \Biggr]
,
\end{equation}
where $p_N^+$ and $p_A^+$ are the momentum of the nucleon and of the nucleus in the light-front frame, respectively, and $A$ the nucleon number of the nucleus ($A=2$ for the deuteron).
We then have $z \le A$.
The masses of the nucleon and of the nucleus are labeled by $m_N$ and $m_A$, respectively.
By nonrelativistically reducing the nuclear binding effect $(p_z^2 + {\mathbf p}^2_\perp)/m_N^2 \ll 1$ and $m_A \approx A m_N$, one obtains \cite{Merabet:1993du}
\begin{equation}
z-1 \approx \frac{p_z}{m_N}
.
\label{eq:pzvsz1}
\end{equation}
This can however be improved by considering the shift of the energy by the moving nucleon inside the deuteron.
The momentum fraction is then
\begin{equation}
z = A \frac{p_N^+}{p_A^+} 
\approx
A\frac{E_N + p_z}{2E_N}
=
1+ \frac{p_z}{E_N}
=
1+\frac{p_z}{\sqrt{p_z^2 + m_N^2}}
,
\end{equation}
where we still neglect ${\mathbf p}_\perp$.
By solving the above equation in term of $p_z$, the nucleon longitudinal momentum inside the deuteron satisfies
\begin{equation}
p_z 
=
\frac{z-1}{\sqrt{z(2-z)}}
m_N
.
\label{eq:pzvsz2}
\end{equation}
We think this manipulation is more suitable for light-front dynamics than the approximation used in Ref. \cite{Merabet:1993du}.

We then apply this variable change to the previously obtained $z$-axis momentum fraction $P(p_z) \equiv |\psi (p_z)|^2$.
We have
\begin{equation}
N_{N / A} (z) dz = \frac{m_N}{\sqrt{z(2-z)}^3} \left| \, \psi \left[ \frac{z-1}{\sqrt{z(2-z)}} m_N \right] \, \right|^2 dz
.
\end{equation}
This relation agrees with the recipe of Ref. \cite{Terentev:1976jk}.

This yields the light-front distribution plotted in \cf{fig:Momentum_fraction_light_front}, 
where one sees that the momentum fraction of the nucleon is broader in the deuteron than in a typical heavy nucleus, which is expected from the importance of the Fermi motion. 

\begin{figure}[htb]
\begin{center}
\includegraphics[width=8.5cm]{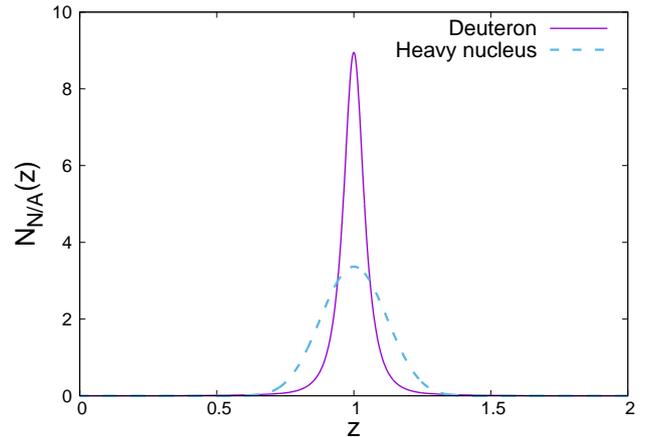}
\caption{\label{fig:Momentum_fraction_light_front}
Momentum fraction of the nucleon in the deuteron. 
The data for a typical heavy nucleus with a Fermi energy $\epsilon_F = 33$ MeV are also shown for comparison (labeled as ``Heavy nucleus'').
}
\end{center}
\end{figure}

\subsection{Gluon distribution\label{sec:gluon_pdf}}

Now that we have the light-front distribution of the nucleon in the deuteron, we can derive the gluon PDF in the deuteron using the impulse approximation, by folding the gluon PDF of the nucleon \cite{Accardi:2016ndt,Dulat:2015mca,Ball:2017nwa,Bonvini:2017ogt,Gao:2017yyd,Lin:2017snn,Alexandrou:2018pbm,Yang:2018bft} by $N_{N/A} (z)$.
Since we are interested in the high-$x$ behavior of the gluon PDF, we need a well behaved gluon PDF up to 1. 
For this reason, we prefer to use GRV98 \cite{Gluck:1998xa}.

The gluon PDF in the deuteron is obtained by folding the gluon PDF of the proton $G^p(x)$ by the light-front distribution of the nucleon inside the deuteron $N_{N/A} (z)$:
\begin{eqnarray}
G^d (x ,\mu_F) 
&=&
2 \int dy dz \, N_{N/A} (z) G^p(y ,\mu_F) \delta (yz -x)
\nonumber\\
&=&
2\int_x^A N_{N/A} (z) \frac{1}{z} G^p(x / z  ,\mu_F) dz
,
\label{eq:convolution_gluon_pdf}
\end{eqnarray}
where $\mu_F$ is the factorization scale.
We note that the effect of the scale evolution is contained in $G^p (x/z,\mu_F)$. 
This operation consists of calculating the contribution depicted by the diagram of \cf{fig:pdf_operator} (a).
In our computation, we of course assume that the proton and the neutron have the same gluon PDF, hence the factor of two in Eq. (\ref{eq:convolution_gluon_pdf}).

\section{Results and discussion\label{sec:result}}

\subsection{Domain of applicability\label{sec:domain}}

Before plotting our results, let us discuss the domain of applicability of our calculation.
Indeed, we assumed that the nucleon inside the deuteron is not modified from the on-shell one.
The nucleons in the deuteron can be considered almost on-shell when the invariant mass of the nucleon pair $M_{pn}$ has a small virtuality compared to the binding of the deuteron:
\begin{equation}
M_{pn}^2 -m_d^2 < m_d \times \epsilon_d
,
\label{eq:virtualitycondition}
\end{equation}
where $m_d$ and $\epsilon_d$ are respectively the deuteron mass and binding energy.
The above condition of virtuality can be converted to a constraint on the nucleon velocity, that is $v = \frac{p_z}{m_N}$ by using Eq. (\ref{eq:pzvsz2}) [or Eq. (\ref{eq:pzvsz1})].
This gives $v < 0.004$ which is obviously nonrelativistic.
From this inequality, we can then derive the corresponding region of the momentum fraction of the gluon in the deuteron, by computing the average $\langle z\rangle$ as a function of $x$.
This yields a conservative limit, $0<x<0.7$, outside which the off-shell correction may be relevant.

Let us now inspect what such off-shell effects may be.
We start by discussing the two-nucleon effects [see \cf{fig:pdf_operator} (b)].
The $n$th moment of the PDF can indeed be expanded in terms of the velocity of the nucleus $v_A$ as \cite{Chen:2004zx}
\begin{equation}
\langle x^n \rangle_{g|A}
=
v_{A, \mu_0} \cdots v_{A, \mu_n}
\langle A| {\cal O}_g^{\mu_0 \cdots \mu_n } |A \rangle
,
\end{equation}
where ${\cal O}_g^{\mu_0 \cdots \mu_n }$ is the gluon density operator.
We note that the nuclear velocity is equal to the nucleon velocity $v$, up to small $x$ corrections due to the nuclear binding.
On the other hand, $\langle x^n \rangle_{g|A}$ can be expressed in terms of the nonrelativistic nucleon operators as
\begin{equation}
\langle x^n \rangle_{g|A}
=
\langle x^n \rangle_{g}
[A +
\langle A| \alpha_n (N^\dagger N)^2 |A \rangle
]
,
\label{eq:pdf_nucleon_correlation}
\end{equation}
where $\langle x^n \rangle_{g}$ is the $n$th moment of the gluon PDF of the nucleon.
The first term $A$ is the nucleon number, obtained from the one-nucleon operator $\langle A| N^\dagger N |A \rangle = A$.
The nuclear matrix element $\langle A| \alpha_n (N^\dagger N)^2 |A \rangle$ provides the nuclear modification effect, and depends on the renormalization scale but not on the momentum fraction.
The coefficient $\alpha_n$ is proportional to the $n$th moment of the nuclear modification effect of the PDF, which is the residual piece of the nuclear PDF after subtracting the gluon PDF of free nucleons.

The zeroth moment $\alpha_0$ is zero, due to charge conservation, and 
the first moment $\alpha_1$ is known to be small from experiment \cite{Rinat:2005qk}.
At the hadron level, the leading off-shell correction is the pion exchange-current \cite{Carbonell:1995yi,Carbonell:1998rj,Chen:2004zx}, but these contributions are N$^3$LO in $\chi$EFT, thus small.
This means that the nuclear modification effect is expected to be small in the nonrelativistic regime.
The first off-shell effect therefore starts from $v^2$ which means that the constraint discussed above, $v< 0.004$, is probably too conservative.

Let us now see the range of velocities in which our framework holds.
In \cf{fig:v2_distribution_gluon}, we plot the averaged squared velocity of the nucleon as a function of the gluon momentum fraction $x$.
We of course exclude the region $\langle v^2 \rangle > 1$ which is unphysical.
We note that $\langle v^2 \rangle $ is still small at $x=1.1$, $\langle v^2 \rangle \approx 0.3$ and therefore 
consider the domain of applicability our our framework as $0 < x< 1.1$, where the off-shell effects are likely small. 

\begin{figure}[htb]
\begin{center}
\includegraphics[width=8.5cm]{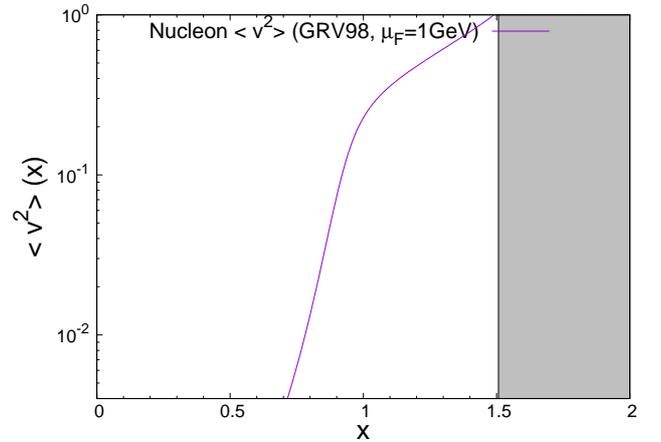}
\caption{\label{fig:v2_distribution_gluon}
The velocity distribution of the nucleon in the deuteron $\langle v^2 \rangle$ as a function of the gluon momentum fraction $x$ obtained in our framework.
The region where $\langle v^2 \rangle > 1$ is unphysical (grey band). 
}
\end{center}
\end{figure}

According to the above discussion, we will show the result of our calculation of the gluon PDF in the deuteron up to $x\simeq 1.1$ in \cf{fig:gluon_pdf}.
The gluon PDF of the deuteron $G^d (x, \mu_F)$ shows a monotonic decrease.
In the region $0 < x < 0.6$, $G^d (x, \mu_F) \approx 2 G^p (x, \mu_F)$ within 5\%, as expected.
It is also notable that the the ratio $G^d / G^p$ is larger than unity for $0< x<0.2$, and that it shows a minimum near $x = 0.4$. 
Above $x \sim 0.6$, the ratio $G^d / G^p$ grows rapidly due to the falloff of the PDF of the proton.
This is due to the Fermi motion, where the momentum of the nucleon in the deuteron is pushed to the high momentum region, in a similar way as the quark PDF.

\begin{figure}[htb]
\begin{center}
\includegraphics[width=8.5cm]{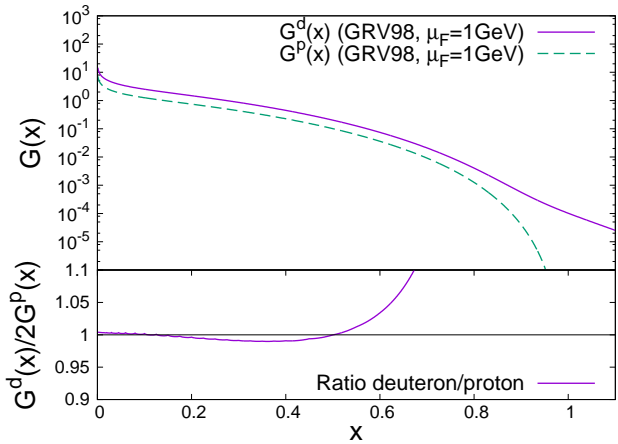}
\caption{\label{fig:gluon_pdf}
Gluon PDF in the deuteron and in the nucleon.
}
\end{center}
\end{figure}

\subsection{Charm distribution of the deuteron\label{sec:charm}}

Another interesting point to discuss is the charm-quark distribution, which can be analyzed in the same way as that of the gluon.
The charm-quark distribution of the deuteron can equally be calculated in the domain of applicability of our framework discussed in Sec. \ref{sec:domain} ($0< x < 1.1$).

The  charm quarks in a nucleon are virtually created by the gluon splitting (see \cf{fig:gluon_splitting}) at leading order.
The distribution of the charm quark generated by this subprocess inherits the gluon distribution, and decreases monotonically in $x$.
We have calculated this contribution by using the charm PDF of CTEQ-JLAB 15 \cite{Accardi:2016qay} which we fold with $N_{N/A}(z)$ discussed in Section \ref{sec:gluon_pdf}.
The result of our calculation is shown in \cf{fig:charm_intrinsic_pdf_comparison}.
The behavior of the charm PDF of the deuteron due to the gluon splitting is similar to that of the gluon.
The ratio of the charm PDFs of the deuteron (per nucleon) to the proton is unity within 5\% for $x<0.4$, and it deviates from unity for $x>0.4$ due to Fermi motion, as expected from the impulse approximation.

\begin{figure}[htb]
\begin{center}
\includegraphics[width=4cm]{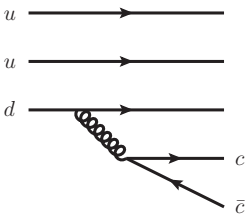}
\caption{\label{fig:gluon_splitting}
Diagrammatic representation of the charm-quark creation in a nucleon via gluon splitting.
}
\end{center}
\end{figure}

\begin{figure}[htb]
\begin{center}
\includegraphics[width=8.5cm]{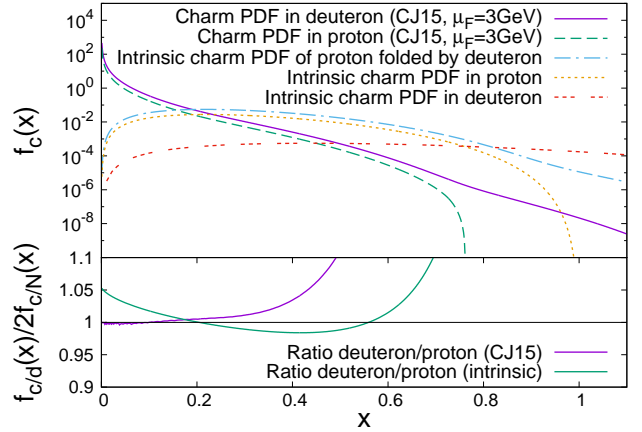}
\caption{\label{fig:charm_intrinsic_pdf_comparison}
Charm PDF in the deuteron and in the nucleon.
}
\end{center}
\end{figure}

The distribution of charm quarks in the nucleon however receives additional non-perturbative contributions from the charm quark-antiquark pair creation which are multi-connected by two or more gluons coupling to different valence quarks (see \cf{fig:intrinsic_charm}).
This {\it intrinsic-charm} contribution, although suppressed since it is higher order in $\alpha_s$, is favored by a higher probability due to the sharing of momenta from different valence quarks. This is in contrast  to the gluon-splitting contributions where the  charm and anti-charm quarks couples to a single valence quark.
In the limit of heavy quarks ($Q$), the intrinsic heavy quark distribution in a hadron is suppressed as $m^{-2}_Q$, as can be derived  
by the application of the operator product expansion~\cite{Brodsky:1984nx,Polyakov:1998rb,Franz:2000ee}.  
A model for the charm distribution in the nucleon based on kinematical constraints is given in Refs.~\cite{Brodsky:1980pb,Brodsky:1981se}
\begin{eqnarray}
f_{c/N}^{\rm int} (x)
&=&
1800 {\cal N} 
x^2 \Bigl[
\frac{1}{3}(1-x)(1+10x+x^2)
\nonumber\\
&& \hspace{5em}
+2x(1+x) \ln x
\Bigr]
.
\label{eq:intrinsic_charm}
\end{eqnarray}
The normalization ${\cal N}$ is phenomenologically determined as ${\cal N} \sim 0.01$ \cite{Brodsky:1981se}.
This distribution peaks at $x\sim 0.2$, and becomes dominant for $x \gtrsim 0.2$

\begin{figure}[htb]
\begin{center}
\includegraphics[width=4cm]{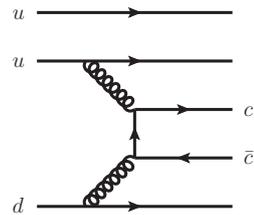}
\caption{\label{fig:intrinsic_charm}
Diagrammatic representation of the intrinsic charm in a nucleon.
}
\end{center}
\end{figure}

We plot in \cf{fig:charm_intrinsic_pdf_comparison} the intrinsic-charm distribution of the deuteron calculated in our framework. 
As is the case of the gluon splitting, the Fermi motion alters the ratio of the deuteron PDFs (per nucleon) to that of the proton from unity for $x>0.6$. 
We also observe that this ratio, although consistent with unity within 5\%, varies more than that of the gluon PDFs in the region $0<x<0.6$.

We can also derive an intrinsic-charm distribution of the deuteron by considering a six-valence-parton configuration (see \cf{fig:intrinsic_charm_6-quark}).
It can be calculated by rescaling the endpoint of Eq. (\ref{eq:intrinsic_charm}) from $x=1$ to $x=2$.
The normalization of the intrinsic-charm contribution to the deuteron is currently not known (we plot it in \cf{fig:charm_intrinsic_pdf_comparison}, with ${\cal N} = 10^{-4}$). There are however some arguments suggesting a sizable contribution of this contribution. Indeed, beside the argument of the momentum-fraction sharing by several valence particles enhancing the intrinsic-charm content
at high $x$, there is another enhancement from the combinatoric factors in the deuteron case.
For the gluon splitting, we obviously have a factor of 6, whereas for the intrinsic charm generated by the radiation of two gluons from two distinct quarks, we have a factor of 15 (see \cf{fig:intrinsic_charm_6-quark} (a)).
The enhancement may even be larger for the intrinsic charm created by the three-gluon emission although it is even higher order in $\alpha_s$, since we have a combinatoric factor of 20 (see \cf{fig:intrinsic_charm_6-quark} (b)).
Note that this combinatoric enhancement is absent in the case of the nucleon.
The intrinsic-charm contribution generated off three-gluon emission may also kinematically be more advantageous than the two-gluon case, since the momenta of valence quarks can stay closer to the valence configuration after the gluon radiation.
It would thus be interesting to perform measurements sensitive to the charm content of the deuteron at $x \sim 1$.
Fixed-target experiments at the LHC with the LHCb or ALICE detector provide an ideal setup for such measurements.

\begin{figure}[htb]
\begin{center}
\includegraphics[width=8.5cm]{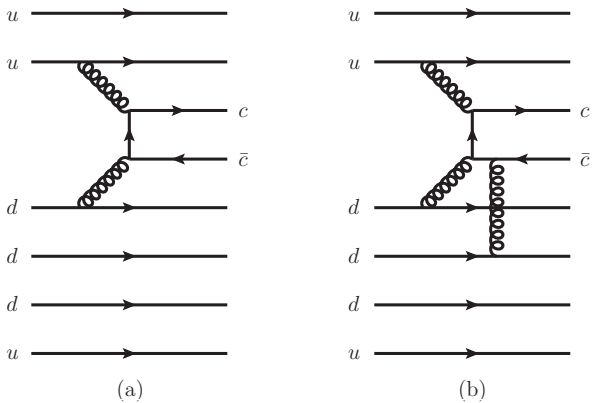}
\caption{\label{fig:intrinsic_charm_6-quark}
Diagrammatic representation of the intrinsic-charm generation in the deuteron: (a) from two-gluon fusion, 
(b) the $\alpha_s$ suppressed --but combinatoric enhanced-- 3-gluon fusion. 
}
\end{center}
\end{figure}

\section{Summary}

In this work, we have calculated the gluon and charm PDFs of the deuteron in the light-front quantization.
We used the impulse approximation where the input nuclear wave function is obtained by solving the nonrelativistic Schr\"{o}dinger equation with the phenomenological Argonne $v18$ nuclear potential as input.
Although we only analyzed the nonrelativistic regime, the range of applicability our computation is estimated to extend up to $x  \sim 1.1$.

We have found that the gluon and charm PDFs of the deuteron (per nucleon) at low $x$ only differs by a few percent from that of the proton, 
as expected for nonrelativistic nucleons in the nucleus.
However as  $x$ becomes close to unity, their distributions deviate significantly from that of the nucleon due to Fermi motion.
This should taken into account when extracting the gluon PDF of the neutron via this system.

We also discussed the charm PDF of the deuteron, which is potentially very interesting at $x\sim 1$ due to the intrinsic-charm contribution.
The intrinsic charm of the deuteron is enhanced by the combinatoric factors characteristic for  gluon emission and the sharing of the momentum by valence partons, although the overall normalization is somewhat uncertain. 
We expect the charm distribution in the deuteron to be studied in the region $0<x<1.1$  by future experiments --particularly in future fixed-target experiments using the LHC beams-- in order to determine the normalization of the intrinsic-charm and hidden-color states.

In the limit of high-momentum scale $Q^2 \to \infty$ for exclusive scatterings, other structures with the same quantum numbers as the $|\, NN \,\rangle$ state, such as the $\Delta \Delta $ states, or the {\it hidden-color configurations}~\cite{Brodsky:1983vf,Brodsky:1985gt,Brodsky:1985gs,Bashkanov:2013cla}, in which quarks are not arranged to form two color-singlet baryons, become relevant as Fock states.
Indeed, in the short distance limit, 80\% of the deuteron will be composed of hidden-color states. 
This state should be continuously be related to the almost maximal $|\, NN \,\rangle$ state at low resolution via the renormalization group equation.
The composition at intermediate momentum scales also involves higher Fock states with a valence gluon~\cite{Hoyer:1997rh}, such as $|\, (uuudddg) \,\rangle$. 
We note that the composition at intermediate distances can only be calculated if the normalization of the Fock state at some scale is known, as is the case for the renormalization group equation analysis. As for now, the implication of these states for inclusive reactions at finite $x$ (away from 2 in the deuteron case), and thus the PDFs, remains to be studied, and is beyond the scope of our exploratory study.

At the endpoint ($x\sim2$), where only one gluon is carrying almost the entire momentum of the deuteron,
the gluon PDF behavior is however related to the form factor of the system at short distances~\cite{Blankenbecler:1974tm,Brodsky:1976mn,Brodsky:1976rz}, and is known analytically.
The counting rules indeed predict $G^d (x) \propto (2-x)^{11}$ \cite{Brodsky:1994kg,Brodsky:1976rz,Brodsky:1976mn,Lepage:1980fj}.
Since the partons are maximally virtual in this limit, the deuteron has to be expressed in terms of quarks and gluons, and it is therefore not possible to discuss with our framework.
Extending our nonrelativistic results to the this limiting case, is also left for a future work especially since it seems difficulty experimentally accessible in a near future.

Our framework could be extended to the case of the gluon and charm PDF in heavier nuclei, such as the $^4$He, which is one of the main ingredient of the interstellar matter, and for $^{14}$N and $^{16}$O, which are the main components of the atmosphere.
Such analyses would be important to reduce the theoretical uncertainty of the cross section of the reactions between primary cosmic rays and the interstellar matter, as well as to predict the ultra-high-energy neutrino background in terrestrial experiments such as IceCube~\cite{Aartsen:2014gkd,Aartsen:2016xlq,Halzen:2016thi,Laha:2016dri,Giannini:2018utr}.
A better knowledge of the gluon PDFs of light nuclei, \eg\ $^3$He and $^4$H, is therefore crucial for high-energy astrophysics, and they could be measured in the near future in LHC fixed-target experiments.

\section*{Acknowledgements} 
We thank Jaume Carbonell and Cedric Lorc\'{e} for useful comments and suggestions.
NY is supported by JSPS Postdoctoral Fellowships for Research Abroad and by the RIKEN iTHES Project.
This work has been supported in part by the French CNRS via the IN2P3 project TMD@NLO and by the French ANR via the LABEX P2IO.
S.J.B. is supported by the Department of Energy, contract DE--AC02--76SF00515. SLAC-PUB--17253.

\bibliographystyle{utphys}

\bibliography{paper_gluon_pdf_in_deuteron}
\appendix

\end{document}